\documentclass[aps,prl,twocolumn,floatfix]{revtex4}

\def\dbarit {{\mathchar'26\mkern-11mud}}
\usepackage{graphicx}
\begin{document}

\title{The second law, Maxwell's daemon and work derivable from
quantum heat engines}

\author{Tien D. Kieu}
\email[]{kieu@swin.edu.au}

\affiliation{Centre for Atom Optics and Ultrafast Spectroscopy,\\
Swinburne University of Technology, Hawthorn 3122, Australia}

\begin{abstract}
With a  class of quantum heat engines which consists of
two-energy-eigenstate systems undergoing, respectively,
quantum adiabatic processes and energy exchanges with heat baths
at different stages of a cycle, we are able to clarify some
important aspects of the second law of thermodynamics.
The quantum heat engines also offer a practical way, as an alternative to Szilard's engine,
to physically realise Maxwell's daemon.  While respecting the second law on the average,
they are also capable of extracting more work from
the heat baths than is otherwise possible in thermal equilibrium.
\end{abstract}
\maketitle Present technology allows for the probing and realisation
of quantum mechanical systems of mesoscopic and even macroscopic
sizes (like those of superconductors, Bose-Einstein condensates,
...), which can also be restricted to a relatively small number of
energy states.  It is thus important to study these quantum systems
directly in relation to the second law of thermodynamics, which has
been applicable to composite systems. We will pursue below this path
without assuming {\em  for the systems} anything extra and beyond
the principles of quantum mechanics~{\footnote{Further materials can
be found in T.D. Kieu, quant-ph/0311157.}}.  Our study is part of a
growing body of investigations into quantum heat
engines~\cite{Zurek:03, Bender:00, Opatrny:02, 
Scully:03, Arnaud:03, entangled}.
Explicitly, the only principles we
will need are those of the Schr\"odinger equation, the Born probability
interpretation of the wavefunctions and the von Neumann measurement
postulate~\cite{vonNeumann:55}.  In particular, we will not exclude, but will make
full use of, any exceptional initial conditions, as long as they are realisable physically.
However, without a better understanding of the emergence of classicality from quantum
mechanics, we will have to assume the thermal equilibrium
Gibbs distributions {\em for the heat baths} which are coupled to the quantum systems.
This assumption is {\em extra} to those of quantum mechanics.

The expectation value of the measured energy of a quantum system is
$ U = \langle E \,\rangle = \sum_i p_i\,E_i$,
in which $E_i$ are the energy levels and $p_i$ are the corresponding
occupation probabilities.  Infinitesimally,
\begin{eqnarray}
dU &=& \sum_i\left(E_i\,dp_i + p_i\, dE_i\right),
\label{first}
\end{eqnarray}
from which we make the following identifications for
infinitesimal heat transferred
$\dbarit Q$ and work done $\dbarit W$,
\begin{eqnarray}
\dbarit Q := \sum_i E_i\, dp_i; \;
\dbarit W := \sum_i p_i\, dE_i.
\label{id}
\end{eqnarray}
Thus, equation~(\ref{first}) is just an expression of the first law,
$dU = \dbarit Q + \dbarit W$.
These identifications
concur with the fact that work done on or by a system can only be
performed through a change
in the generalised coordinates of the system, which in turn gives rise to a
change in the distribution of the energy levels~\cite{Schrodinger:89}.

Our quantum heat engines are just two-level quantum systems,
which are the quantum version of the Otto engines~\cite{Arnaud:03}.
(They are readily extendable to
systems of many discrete energy levels.)  They could be realised with coherent
macroscopic quantum systems like, for instance, a Bose-Einstein condensate confined
to the bottom two energy levels of a trapping potential.
The exact cyclicity will be enforced to ensure that upon completing
a cycle all the output products of the engines are clearly displayed without any hidden effect.

A cycle of the quantum heat engine has four stages:
\begin{enumerate}
\item {\em Stage 1:}  The system has the probability $1 - p_u^{\,(1)}(0)$
in the lower state {\em prior} to some kind of contact
with a heat bath at temperature $T_1$.
After some contact time $\tau_1$, the system achieves a probability
$p_u^{\,(1)}(\tau_1)$ to gain some energy from the heat bath to jump up an energy
gap of $\Delta_1$ to be in the upper state.  Only heat is transferred in this stage
to yield a change in the occupation probabilities.
\item {\em Stage 2:}  The system is then isolated from the heat bath and undergoes
a quantum adiabatic
expansion to reduce the energy gap from $\Delta_1$ to a smaller
value $\Delta_2$.  The probability in the upper state is
maintained throughout (provided the expansion rate
is sufficiently slow according to the quantum adiabatic theorem~\cite{Messiah}),
an amount of work is thus performed {\em by} the system, but no heat is
transferred.\item {\em Stage 3:}  The system, with a probability
$p_u^{\,(2)}(0)$ being in the upper state, is brought into some kind of contact
with another heat bath at temperature $T_2$ for some time $\tau_2$ until it
gains a probability $1-p_u^{\,(2)}(\tau_2)$ to release some energy to the bath
and jump down the gap $\Delta_2$ to be in the lower state.
\item {\em Stage 4:}  The system is removed from the heat bath
and undergoes a quantum adiabatic contraction to increase the energy
gap from $\Delta_2$ back to the larger value $\Delta_1$.
An amount of work is performed {\em on} the system in this stage.
\end{enumerate}

Note that we need not and have not assigned any temperature to the quantum system; all
the temperatures are properties of the heat baths, which are assumed to be in
the Gibbs state.
However, in the operation above, the absorption and release of energy in
stages 1 and 3 occur neither definitely, nor even deterministically.
Quantum mechanics tells us that they can only happen probabilistically; and
the probabilities that such transitions take place depend on the details of the
interactions with and some intrinsic properties (namely, the temperatures)
of the heat baths.
The cyclicity of the heat engines then puts a constraint on the probabilities,
$p_u^{\,(1)}(0) = p_u^{\,(2)}(\tau_2)$, and
$p_u^{\,(2)}(0) = p_u^{\,(1)}(\tau_1)$.
The net work done by our quantum heat engines in the two quantum adiabatic
passagesin stages 2 and 4 is, from~(\ref{id}),  given that $\Delta_1>\Delta_2$,
\begin{eqnarray}
\Delta W
&=& \left({p}_u^{\,(1)}(\tau_1)-{p}_u^{\,(2)}(\tau_2)
\right)\left(\Delta_2-\Delta_1\right).
\label{5}
\end{eqnarray}

If the system is allowed to thermalise with the heat baths in stages 1 and 3,
the thermal equilibrium probabilities, for $i=1,2$,
\begin{eqnarray}
\tilde{p}_u^{\,(i)} &=& 1/\left(1 + \exp({\Delta_i}/{kT_i})\right),
\label{thermal}
\end{eqnarray}
are to be used in~(\ref{5}).
These probabilities are definitely {\em non-zero}.
Thus even in the case of ($T_1\le T_2$)
there {\em exist}, even with a small probability
$\tilde{p}_u^{\,(1)}(1-\tilde{p}_u^{\,(2)})$, cycles in which the transitions described
in stages 1 and 3 above actually take place.   (The probability, however, diminishes
exponentially for $n_c$ consecutive cycles, $(\tilde{p}_u^{\,(1)}
(1-\tilde{p}_u^{\,(2)}))^{n_c}$,
in all of which the system performs net work on the environment.)
As a result, there {\em are} certain cycles {\em whose \underline{sole} result is the
absorption of heat from a reservoir and the conversion of this heat into work,
of the amount $(\Delta_1-\Delta_2)$}.
This amounts to a violation of the Kelvin-Planck statement of the second law due to the
explicit probabilistic nature of quantum mechanical processes.  This violation,
however, occurs only randomly, with some small probability, and thus is not
exploitable.
%

However, there exists a sure way to {\em always} extract work, to the amount of
$(\Delta_1 - \Delta_2)$,
in each {\em completable} ``cycle."  In order to eliminate the
probabilistic uncertainty,
we prepare the system to be in the lower state prior to stage 1;
make an energy measurement after stage 1 and only let the engine
continue to stage 2 {\em if} the measurement result confirms that the system is
in the upper state; we then make another measurement after stage 3 and only let
the engine continue to stage 4 {\em if} the measurement result confirms that the
system is in the lower state.  All this can be carried out even for the case
$T_1\le T_2$ to extract, in a controllable manner, some work which would have
been otherwise prohibited by the second law.  This apparent violation of the
second law is analogous to that of Szilard's one-atom engine and is nothing but
the result of an act of Maxwell's daemon~\cite{Demon:90}.  Our quantum heat
engines can thus provide a feasible and quantum way to realise Maxwell's daemon
in a different way to Szilard's engine.  Indeed, the condition of strict
cyclicity of each engine's cycle is broken here.  After each cycle
the measurement apparatus, being a Maxwell's daemon, has already registered the
results which are needed to determine the {\em next} stages of the engine's
operation.  In this way, there are extra effects and changes to the
register/memory of the apparatus, even if we assume that the quantum measurement
steps themselves cost no energy and leave no net effect anywhere else.  To remove
these remnants in order to restore the strict cyclicity, we would need to bring
the register back to its initial condition by erasing any information obtained
in the cycles, either by resetting its bits or by thermalising the register with
some heat bath.   Any which way, extra effects are inevitable, namely an amount
of heat of at least $kT\ln 2$ will be released per bit erased.  The second law
is thus saved~\cite{Lloyd:97}.  More extensive discussions and debates on these
issues can be found in the literature~\cite{Demon:90}.%

Despite the random quantum
mechanical violation, the second law is upheld on the average.
The work that can be performed {\em by} our engines in thermal equilibrium when
and only when,
as can be seen from~(\ref{5}), ${p}_u^{\,(1)}(\tau_1)>{p}_u^{\,(2)}(\tau_2)$
and thus when and only when, from~(\ref{thermal}),
\begin{eqnarray}
{T_1} &>& T_2\left({\Delta_1}/{\Delta_2}\right).
\label{6}
\end{eqnarray}
The expressions~(\ref{5}) and~(\ref{6}) (which was also derived in a more
specific context in~\cite{Feldmann:2000}) not only confirm the broad validity
of the second law but also refine the law further in specifying how much $T_1$
needs to be larger than $T_2$ before some work can be extracted.  In other
words,{\em work cannot be extracted, on the average, even when $T_1$ is greater
than $T_2$ but less than $T_2(\Delta_1/\Delta_2)$}, in contradistinction to the
classical necessary condition that $T_1$ only needs to be larger than $T_2$.
(However, for Carnot's engines, see~\cite{Arnaud:03}, the ratio of successive energy gaps
involved tends to one.)

The efficiency of our engines is~{\footnote{A similar expression was also
obtained through a specific context in~\cite{Scovil:59}.}}
\begin{eqnarray}
\eta_q &=& {\Delta W}/{Q_{\rm in}} = \left(1-{\Delta_2}/
{\Delta_1}\right),
\label{eff}
\end{eqnarray}
which is {\em independent} of heat bath temperatures and is the maximum available
within the law of quantum mechanics.  This also serves as the upper bound, with
appropriate $\Delta_1$ and $\Delta_2$, of the efficiency of any heat engine,
including Carnot's engines, because the work performed by our heat engines
through their quantum adiabatic processes is the maximum that can be
extracted~\cite{Landau:80}.  The efficiency of Carnot's engines, $\eta_C = 1 -
T_2/T_1$, can be derived from~(\ref{eff}) through an infinite number of infinitesimal quantum
adiabatic processes~\cite{Arnaud:03}.

We present next a modification of
the quantum heat engines which can better the work derivation than that
available from thermal equilibrium.
A modified cycle also has four stages, of which stage 2 and stage 4 remain the same
as described previously, whereas stages 1 and 3 are
replaced respectively by:
\begin{enumerate}
\item {\em Stage 1':}  The system is entered to a single-mode cavity which is tuned to
match the energy gap $\Delta_1$ of
the system and which is in thermal equilibrium with a heat bath at temperature $T_1$.
The thermal distribution for the cavity mode is given by~(\ref{thermaldist}) below.
After some time interval
$\tau_1$, the system is removed from the cavity to enter stage 2.
\item {\em Stage 3':}
Like stage 1' above but with temperature $T_2$ and smaller energy gap $\Delta_2$.
After some time $\tau_2$, the system departs the cavity for stage 4.
\end{enumerate}

In each of the cavities so described, the state of the system satisfies the Schr\"odinger equation for a
single two-level atom interacting with a single-mode field which has the
thermal probability $P_n(T_i)$ to find exactly $n$ photons of frequency $\Delta_i/h$
in the cavity at temperature $T_i$,
\begin{eqnarray}
P_n(T_i) &=& \frac{1}{1+\bar{n}_i}\left(\frac{\bar{n}_i}{1+\bar{n}_i}\right)^n,
\label{thermaldist}
\end{eqnarray}
where $\bar{n}_i$ is the thermal average boson number,
$
\bar{n}_i = 1/\left({e^{{\Delta_i}/{kT_i}}-1}\right)\, .
\label{planck}
$
The {\em exact} solution of this Schr\"odinger equation is given in~\cite{Scully:97},
from which we arrive at 
\begin{eqnarray}
p_u^{\,(i)}(t) &=& \frac{(1+2\bar{n}_i)\,p_u^{\,(i)}(0)
- \bar{n}_i}{(1+\bar{n}_i)}\sum_{n=0}^\infty P_n(T_i)
\cos^2\left(\Omega_n t\right)\nonumber\\
&& + \frac{\bar{n}_i( 1 - p_u^{\,(i)}(0))}{1+\bar{n}_i},
\label{upperprobability}
\end{eqnarray}
where $\Omega_n = g\sqrt{n+1}$ and $g$ is the coupling constant between the system
and the cavity mode.  
From~(\ref{upperprobability}) we can derive the bounds of the probability
$p_u^{\,(i)}(t)$ in terms of the initial probability
and temperature,
\begin{eqnarray}
p_u^{\,(i)}(t) &\in& [p_u^{\,(i)}(0),
{\bar{n}_i( 1 - p_u^{\,(i)}(0))}/{(1+\bar{n}_i)}], \nonumber\\
&& {\mbox{\rm for }} p_u^{\,(i)}(0)\in [0,\bar{n}_i/(1+2\bar{n}_i)];\nonumber\\
p_u^{\,(i)}(t) &\in& [{\bar{n}_i( 1 - p_u^{\,(i)}(0))}/{(1+\bar{n}_i)},
p_u^{\,(i)}(0)], \nonumber\\
&& {\mbox{\rm for }} p_u^{\,(i)}(0) \in [\bar{n}_i/(1+2\bar{n}_i), 1].
\label{9}
\end{eqnarray}

\begin{figure}
\includegraphics[scale=0.4]{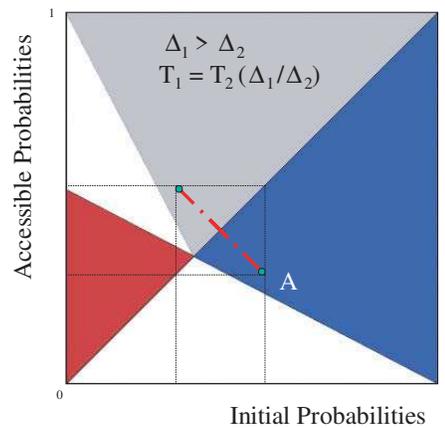}%
\caption{\label{fig1}For the cavity in contact with
a heat sink at $T_2$, 
the accessible probability at all times, given a particular initial probability,
is bounded in the (blue) area under the diagonal, with its reflection shown in
gray above the diagonal.  The bounded (red) area above the diagonal is for the
cavity in contact with a heat bath at $T_1 = T_2(\Delta_1/\Delta_2)$.}
\end{figure}

Figure~\ref{fig1} depicts these bounds.
The probability at any subsequent time is only accessible in the bounded, dark
triangular areas, each of which corresponds to a different temperature, on the
two sides of the diagonal.
For the cavity in contact with the heat bath at $T_1$,
we want to have the exit probability to be greater (the greater, the better) than the initial
probability,
$
p_u^{\,(1)}(\tau_1) > p_u^{\,(1)}(0) = p_u^{\,(2)}(\tau_2),
$
thus we need only to consider the lower area for this
temperature.  Reversely, for the cavity in contact with the heat bath at $T_2$,
we want to have the exit probability to be smaller (the smaller, the better) than
the initial probability, $
p_u^{\,(2)}(\tau_2) < p_u^{\,(2)}(0) = p_u^{\,(1)}(\tau_1),
$
hence we need only to consider the upper area for this temperature.
Thus, Figure~\ref{fig1} combines the two heat
baths in which the upper area comes from
$T_1$ and the lower from $T_2$, with $T_1 = T_2(\Delta_1/\Delta_2)$.
The coordinate of a point $A$ in the lower area of Fig.~\ref{fig1} is
$(p_u^{\,(2)}(0), p_u^{\,(2)}(\tau_2))$, representing an exit probability less than
the entry one at the cavity with temperature $T_2$.  The corresponding point at temperature
$T_1$ must have, by requirement of cyclicity, the coordinate
$
(p_u^{\,(1)}(0), p_u^{\,(1)}(\tau_1)) = (p_u^{\,(2)}(\tau_2), p_u^{\,(2)}(0)),
$
which thus is the reflection of the point $A$ across the diagonal.  However, it is clearly seen
that for $T_1 \le T_2(\Delta_1/\Delta_2)$ this reflection is {\em not} in the accessible
upper area.  We then conclude that at these temperatures, even with $T_1$
greater than $T_2$ by a factor $(\Delta_1/\Delta_2)$, the quantum heat engines
{\em cannot} do work {\em on the average}.
Thus, we have once again confirmed the second law that,
{\em on the average}, no process is possible whose \underline{sole} result is the
transfer of heat from a cooler to a hotter body, with or without
a production of work.
But we have also clarified the degrees of coolness and hotness in terms of the quantum
energy gaps involved before such a process is possbile; namely, we must have, as
a {\em necessary} but not sufficient condition, $T_1 > T_2(\Delta_1/\Delta_2)$
as in~(\ref{6}).

The perfect agreement between this specific result derived from the quantum dynamical bounds~(\ref{9})
with the general result~(\ref{6}) derived from statistical mechanics is quite remarkable -- but
is yet to be fully understood and hence deserved further investigation elsewhere.
We speculate that such agreement is not accidental but is a consequence of the Gibbs
distributions assumed for the heat baths in both derivations;
and it should thus be independent of specific details of the quantum dynamics involved.

\begin{figure}
\includegraphics[scale=0.4]{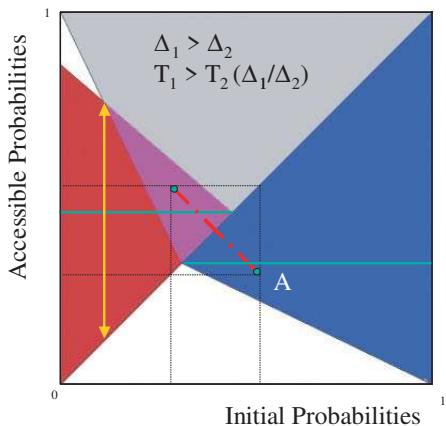}%
\caption{\label{fig3}Similar to Fig.~\ref{fig1} but this time with $T_1>
T_2(\Delta_1/\Delta_2) > T_2$.
The maximum work can be extracted in a single cycle is proportional to the length
of the (yellow) vertical double arrow.} 
\end{figure}

Fig.~\ref{fig3} combines the cases $T_2$ and
$T_1>T_2 (\Delta_1/\Delta_2)$ in which there is some overlap between the areas above the diagonal and thus makes possible the
production of some work, $\Delta W_{\rm cav}$.
If and when we choose to operate with a point below the thermal equilibrium
line in the blue area (for $T_2$) such that its reflection across the diagonal is above the
thermal equilibrium line in the red area (for $T_1$) as shown
in the figure, we can derive more work than the case of thermal equilibrium,
$\Delta W_{\rm th}$.  This is because
the net work done is proportional to the difference in probabilities as shown
in~(\ref{5}).  Here $\left|\Delta W_{\rm cav}\right|$ is greater than
$\left|\Delta W_{\rm th}\right|$
because the vertical distance between point $A$ and its reflection in
Fig.~\ref{fig3} is greater than the vertical distance between the two horizontal lines,
which represent the two thermal equilibria.

In summary, by interpreting work and heat, but without introducing entropy directly, in the
quantum domain and by applying this interpretation to the simplest quantum
systems of two levels, we have not only confirmed the broad validity of the
second law but also been able to clarify and refine its various aspects.
On the one hand, the second law is seen to be valid {\em on the average}
in the broad sense.
On the other hand, explicitly because of the probabilistic nature of quantum mechanics,
there do exist physical processes which can violate certain classical statements of the
second law.  (Such violation, while can be observed in some cycles of our
two-level heat engine, may not be
observable in any single cycle for a system which consists of many subsystems,
because the averaging effects over the subsystems already occur in {\em each} cycle
for the composite system.)
However, such violation only occurs randomly and can thus
neither be exploitable nor harnessible.
This confirmation of the second law is in accordance
with the fact that, while we can treat the quantum heat engines purely and entirely
as quantum mechanical systems, we still have to assume the Gibbs distributions {\em for the
heat baths} involved.
Such distributions can only be derived~\cite{Landau:80} with non-quantum-mechanical
assumptions, which ignore any quantum entanglement within the heat baths.  Indeed,
it has been shown that~\cite{Lenard:78, Tasaki:00} the law of entropy increase is a
mathematical consequence of the initial states being in such general
equilibrium distributions.  This illustrates and highlights the connection between the
second law to the unsolved problems of emergence of classicality, of quantum
measurement and of decoherence, all of which are central to quantum mechanics.
Only until some further progress can be made on these problems, the classicality of the heat
baths will have to be assumed and remained in the assumption of the
Maxwell-Boltzmann-Gibbs thermal equilibrium
distributions.

We have also clarified the degree of temperature
difference~(\ref{6}), in terms of the quantum energy gaps involved, between the
heat baths before any work can be extracted.  We have also shown how to extract
more work from the heat baths than otherwise possible with thermal equilibrium.
Note that such an operation is subject to the bounds given in~(\ref{9}), which then,
as can be seen through their depiction in the figures, ensure that we stay within the second law.
Note also that this
is not an operation of Maxwell's daemon because the information about the time intervals
$\tau_1$ and $\tau_2$ is fixed and forms an integrated part of the modified
engines.  This information, being common to {\em all} cycles,
need not and  should not be erased after each cycle.
Finally, our class of quantum heat engines can readily
offer a feasible way to physically realise Maxwell's daemon, in a way different
to Szilard's engine but also through the acts of quantum
measurement and information erasure.

\begin{acknowledgments}
I would like to acknowledge Alain Aspect for his seminar
which directly prompted this investigation.
I would also like to thank Bryan Dalton, Jean Dalibard, Peter Hannaford, Peter Knight
and Bruce McKellar for helpful discussions; Jacques Arnaud, Michael Nielsen, Bill Wootters
and Wojciech Zurek for email correspondence.
\end{acknowledgments}

\bibliography{qhengine}

\end{document}